\documentclass[
    final            
  ]
  {aipproc}

\layoutstyle{6x9}

\begin{document}

\title{Optical absorption in small BN and C nanotubes}

\author{L. Wirtz}{
address={Department of Material Physics, University of the Basque
Country, Centro Mixto CSIC-UPV, and Donostia International Physics Center,
Po.~Manuel de Lardizabal 4, 20018 San Sebasti\'an, Spain}
}

\author{V. Olevano}{
address={Laboratoire des Solides Irradi\'es, UMR 7642 CNRS/CEA,
Ecole Polytechnique, F-91128 Palaiseau, France}
}

\author{A. G. Marinopoulos}{
address={Laboratoire des Solides Irradi\'es, UMR 7642 CNRS/CEA,
Ecole Polytechnique, F-91128 Palaiseau, France}
}

\author{L. Reining}{
address={Laboratoire des Solides Irradi\'es, UMR 7642 CNRS/CEA,
Ecole Polytechnique, F-91128 Palaiseau, France}
}

\author{A. Rubio}{
address={Department of Material Physics, University of the Basque
Country, Centro Mixto CSIC-UPV, and Donostia International Physics Center,
Po.~Manuel de Lardizabal 4, 20018 San Sebasti\'an, Spain}
}

\begin{abstract}
We present a theoretical study of the optical absorption spectrum
of small boron-nitride and carbon nanotubes using
time-dependent density-functional theory and the random phase approximation.
Both for C and BN tubes, the absorption of light polarized perpendicular
to the tube-axis is strongly suppressed due to local field effects.
Since BN-tubes are wide band-gap insulators, they only absorb
in the ultra-violet energy regime, independently of chirality and
diameter. In comparison with the spectra of the single C and BN-sheets,
the tubes display additional fine-structure which 
stems from the (quasi-) one-dimensionality
of the tubes and sensitively depends on the chirality and tube diameter.
This fine structure can provide additional information for the assignment 
of tube indices in high resolution optical absorption spectroscopy.
\end{abstract}

\maketitle


Just as a carbon nanotube can be thought of as a rolled up
graphene sheet, a hexagonal single sheet of BN can be used to
construct a BN nanotube. These tubes are isoelectronic to carbon 
tubes, but carry over some of the characteristic differences
of hexagonal BN with respect to graphite. In particular,
BN-tubes have a bandgap similar to h-BN, mostly independent
of the tube diameter and chirality \cite{rubi94,blas94}.
Related to this large band gap (the DFT-band gap is 4 eV (see Fig.~1) while
the quasi-particle band gap in the GW approximation amounts to 5.5 eV
\cite{blas94}) one expects a high thermal stability and relative
chemical inertness for BN-tubes as compared to its carbon counterparts.

After first synthesis of multi-wall BN-tubes was reported in 1995 
\cite{chop95}, 
multi and single wall BN-tubes are now routinely produced in several 
groups, the latest success being the production of single-wall BN-tubes in 
gram quantities \cite{lee01}. The challenge now consists 
in the spectroscopical characterization of nanotube samples (both C and BN)
and, if possible, single isolated nanotubes. 
The final goal is to find a unique mapping of the measured
electronic and vibrational properties onto the tube indices $(n,m)$.
One possible spectroscopic method is optical absorption
spectroscopy where direct excitation from occupied to unoccupied
states leads to photon absorption.

The energy difference $E_{ii}$ between corresponding occupied and unoccupied 
Van Hove singularities (VHSs) in the 1-dimensional electronic density of 
states (DOS) of C-nanotubes is approximately inversely proportional to 
the tube diameter $d$. In resonant Raman spectroscopy and scanning
tunneling spectroscopy, this scaling is employed for the determination of tube 
diameters. A recent, spectacular example is
the fluorescence spectroscopy of single carbon tubes in aqueous solution,
where $E_{22}$ is probed through the frequency of the excitation laser
and $E_{11}$ is probed simultaneously through the frequency of the emitted
fluorescent light \cite{bac03}.
For the distance between the first VHSs in
semiconducting C-tubes, a simple $\pi$-electron
tight-binding fit yields the relation
$E_{11}=E_{22}/2=2a_{C-C} \gamma_0/d$, where $a_{C-C}$ is the 
distance between nearest neighbor carbons.
The value for the hopping matrix element $\gamma_0$
varies between 2.4 eV and 2.9 eV, depending on the experimental context
in which it is used. This fact is a clear indication that
the above relation gives only qualitative and not quantitative information
on the tube diameter. Furthermore, for small
tubes the band structure completely changes with respect to the band
structure of large diameter tubes, including a reordering of the VHSs in
the density of states and displaying fine structure beyond the first 
and second VHSs (Fig. 1). This structure sensitively depends on the tube 
indices and may be probed by optical absorption spectroscopy over a wider 
energy range (possibly extending into the UV regime).

The scope of this paper is to use {\it ab initio} techniques to
go beyond the tight-binding estimate for excitation energies and to 
uncover some of these additional features present in the optical spectra 
of small C and BN nanotubes. In order to develop an intuitive understanding
of the optical absorption in tubes, we compare 
with the absorption in single sheets of graphene and h-BN. 
In this paper we present results for the very small
C(3,3) and BN(3,3) tubes while work for larger tubes is in progress.
C(3,3) tubes seem not to exist as free single-wall 
tubes, but they have been grown and shown to be stable inside the cylindrical
channels of a zeolite crystal \cite{li01}. The band structure and 
density of states (DOS) of the two tubes are shown in Fig.~1. 
The following calculations will show which vertical excitations from unoccupied 
to occupied states yield the dominant contributions to the absorption spectra.

\begin{figure}
  \includegraphics[width=.9\textwidth]{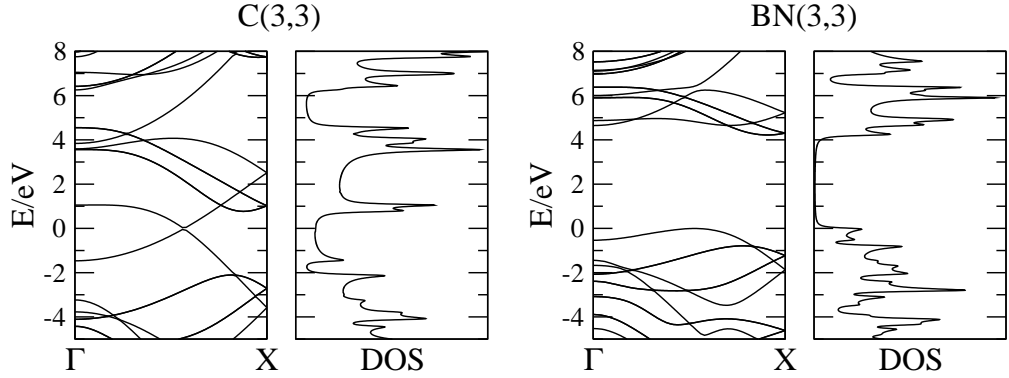}
  \caption{Electronic band structure and density of states (DOS)
of a C(3,3) nanotube (left side) and a BN(3,3) nanotube (right side). 
Zero energy denotes the upper edge of the valence band. The calculation
has been performed using DFT in the local density approximation.}
\end{figure}

The cross section $\sigma(\omega)$ for optical absorption at frequency $\omega$ 
is proportional to the imaginary part of
the macroscopic dielectric response function of the system \cite{noteps}: 
$\sigma(\omega) \propto \mbox{Im}(\epsilon_M(\omega))$.
We evaluate $\epsilon_M$ using linear response
theory \cite{lrdft} within the general framework of time-dependent
density-functional theory (TDDFT).

First, we determine the ground state equilibrium geometry of the system
using the code ABINIT \cite{abinit,params}. From
the ground state density we compute the 
one-particle states $|n,{\bf k}\rangle$ and energies $\epsilon_{n,{\bf k}}$
(labeled by Bloch wave vector ${\bf k}$ and band index $n$) 
for all occupied and a large set of unoccupied bands. 

\begin{figure}
  \includegraphics[width=.92\textwidth]{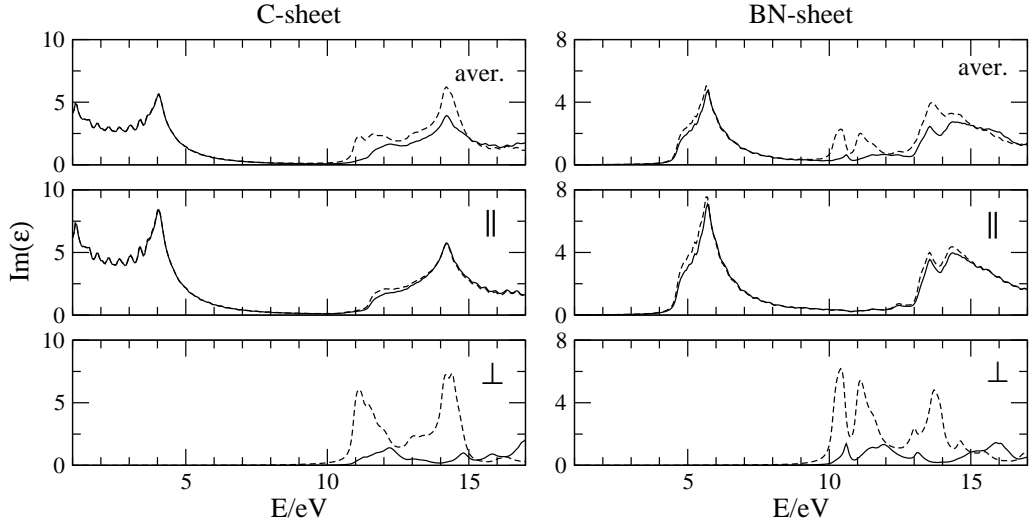}
  \caption{Imaginary part of the computed dielectric function (arb. units) 
for a graphene sheet (left side) and a single hexagonal BN-sheet (right side). 
Lower panels: light polarization perpendicular to the plane; middle panels: 
polarization parallel 
to the plane; upper panels: spatial average. Solid lines: calculation
with local field effects; dashed lines: without LFE (see text for details).}
\end{figure}

The next step is the calculation of the independent particle polarizability
$\chi^0$ \cite{notechi}. It involves a sum over excitations
from occupied bands to unoccupied bands \cite{dp}:
\begin{eqnarray}
\chi^0_{{\bf G},{\bf G'}}({\bf q},\omega) =  \ \ \ \ \ \ \ \ \ \nonumber
\ \ \ \ \ \ \ \ \ \ \ \ \ \ \ \ \ \ \ \ \ \ \ \ \ \ \ \ \ \ \ \ \ \ \ \ \ \ \ \
\ \ \ \ \ \ \ \ \ \ \ \ \ \ \ \ \ \ \ \ \ \ \ \ \ \ \ \ \ \ \ \ \ \ \ \ \ \ \ \
\ \ \ \ \ \ \ \ \ \ \ \ \ \ \ \ \ \ \ \ \
  \\
2 \int \frac{d k^3}{(2\pi)^3} \sum_n^{occ.} \sum_m^{unocc.}
\left[\frac{\langle n,{\bf k}|e^{-i({\bf q}+{\bf G}){\bf r}}|m,
                                                   {\bf k}+{\bf q}\rangle
     \langle m,{\bf k}+{\bf q}|e^{i({\bf q}+{\bf G'}){\bf r'}}|n,{\bf k}\rangle}
     {\epsilon_{n,{\bf k}} - \epsilon_{m,{\bf k}+{\bf q}} - \omega \ - i\eta}
    - (m \leftrightarrow n) \right],
\label{chi0}
\end{eqnarray}
where $(m \leftrightarrow n)$ means that the indices $m$ and $n$ of the
first term are exchanged. 
The result is checked for convergence with respect to the number of bands 
\cite{nvalues} and the discrete sampling of $k$-points within the first 
Brillouin zone \cite{kvalues}.
Using the random phase approximation \cite{lrdft}, 
the ``longitudinal''
dielectric function is obtained through 
$\epsilon_{{\bf G},{\bf G'}}({\bf q},\omega) = 
1 - V_c({\bf q}+{\bf G}) \chi^0_{{\bf G},{\bf G'}}({\bf q},\omega)$,
where $V_c({\bf q}) = 4\pi/|{\bf q}|^2$ is the Coulomb 
potential in reciprocal space.
Finally, the macroscopic dielectric response is given by
$\epsilon_M(\omega) = 1/\epsilon^{-1}_{00}({\bf q} \rightarrow 0,\omega)$.
The limit ${\bf q} \rightarrow 0$ depends on the 
direction of ${\bf q}$, i.e., the polarization of the electric field.
The difference between $\epsilon_{00}$ and $\epsilon_M$ is due to the
inhomogeneity of the response of the system and is called 
``local field effects'' (LFE). For convergence of $\epsilon_M$, 
$\epsilon_{{\bf G},{\bf G'}}$ has to be calculated for a sufficient number 
of ${\bf G}$-vectors.
The matrix is then inverted and $\epsilon_M$ is obtained from the head
of the inverse matrix as $(\epsilon^{-1}_{00})^{-1}$. 

The optical absorption spectra of a single graphene sheet and a sheet of
h-BN are displayed in Fig.~2 (extended far into the region of UV light). 
The spectra are strongly dependent on the 
polarization of the laser beam. The main difference between C and BN
can be seen for the polarization parallel to the plane: While the C-sheet
absorbs for all frequencies in the visible light range (the ``color'' of 
graphite is black), absorption in BN only sets in above 4 eV, i.e. in 
the region of UV light (since DFT underestimates the band-gap, we
expect a blue-shift of the onset by $\approx$ 1.5 eV). The high frequency 
part of the spectra are quite 
similar because C and BN are isoelectronic and the high-lying unoccupied 
states are less sensitive to the difference in the nuclear charges than the 
states at and below the Fermi energy. Local field effects turn out
to be unimportant for parallel polarization.
The absorption spectra in perpendicular polarization
are remarkably similar for C and BN: up to 9 eV both sheets
are completely transparent. Furthermore, in both cases, LFE
lead to a strong reduction of absorption at energies higher
than 9 eV (redistribution of oscillator strength to even higher energies).
\begin{figure}
  \includegraphics[width=.92\textwidth]{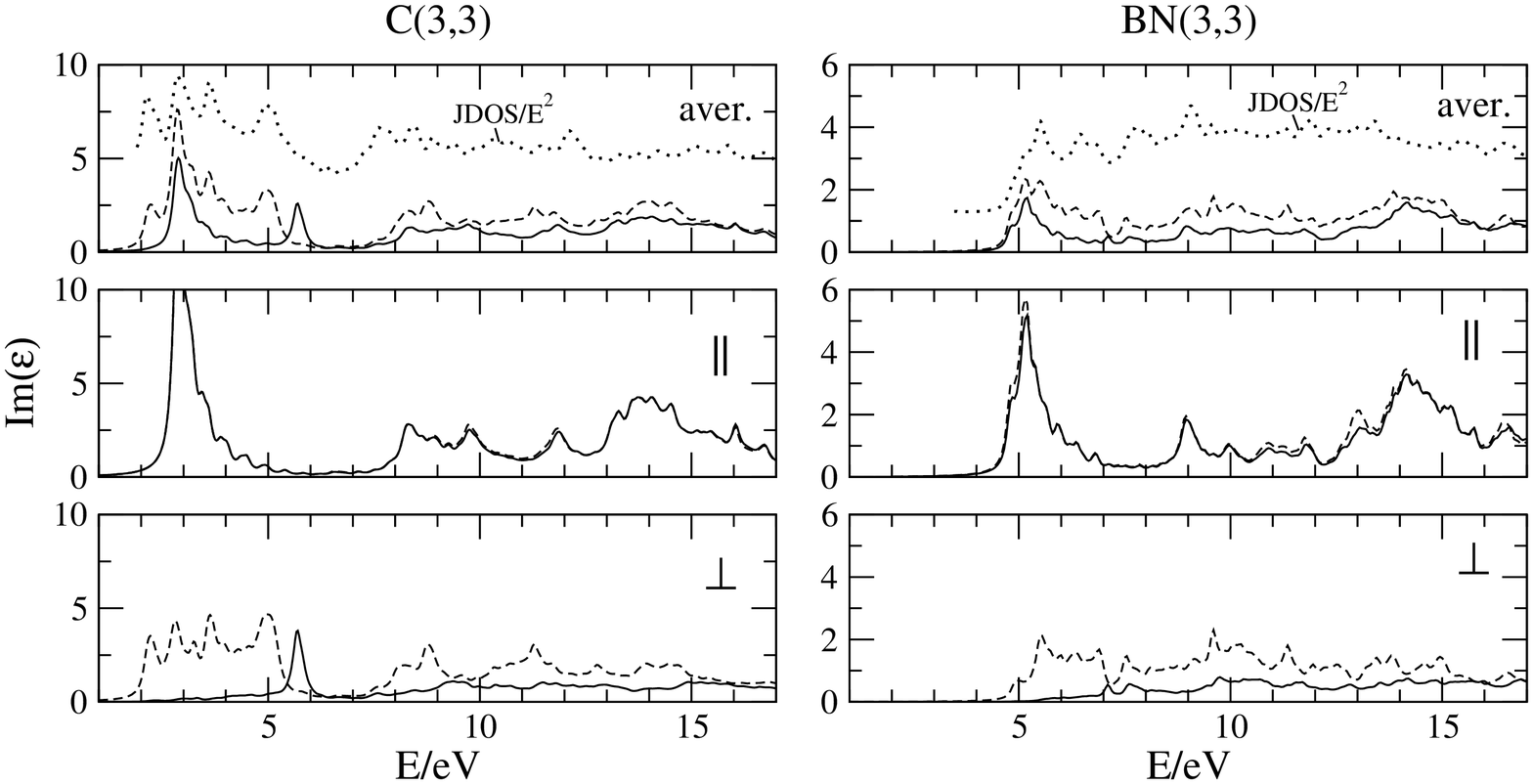}
  \caption{Imaginary part of the dielectric function (arb. units) 
for a C(3,3) tube (left side) and BN(3,3) tube (right side). Lower panels: 
light polarization perpendicular to the tube; middle panels: polarization
 parallel to the tube axis; upper panels: spatial average. Solid lines: 
calculation with local field
effects; dashed lines: without LFE. The joint
density of states (divided by $E^2$) is indicated by dotted lines.}
\end{figure}

Fig.~3 displays the absorption spectra for the (3,3) tubes of C and BN.
For comparison, in the panel of the spatially averaged spectra,
we have also included the joint density of states (JDOS), divided
by the square of the transition energy. If LFE are neglected, most peaks of 
the JDOS are visible in the averaged absorption spectra while some
peaks are suppressed due to small or vanishing oscillator strength in 
Eq.~(\ref{chi0}).
Proper inclusion of LFE leads to a smoothing of the spectra and to a
redistribution of oscillator strength to higher energies for polarization
perpendicular to the tube axis. However,
some fine structure survives and may be discernible in high-resolution
optical absorption experiments. This fine structure is not an 
artifact of low k-point sampling but is due to the presence of VHSs in the 
1-dim.\ DOS of the tubes (Fig.~1).
In the absorption spectrum polarized along the axis of the C(3,3) tube, the 
pronounced peak at 3 eV corresponds to the first (dipole-allowed) transition
between VHSs in Fig.~1. Even though the tube is metallic, absorption
at lower energies is suppressed due to the dipole-selection rules.
The position of the first absorption peak in C-tubes not only depends
on the tube radius but also (for fixed radius) on the index pair $(n,m)$
and varies up to 2 eV for very small tubes\cite{li01,liu02,mach02,tolis03}. 
In contrast, the onset of absorption in BN-tubes 
corresponds to the threshold in the BN-sheet and is
mostly independent of the tube indices (only the peak structure
above the onset varies with the tube indices). Between 7 and 12 eV,
both C and BN tubes display a similar pattern of peaks which are
absent in the corresponding sheets.
As in the case of the sheets, LFE are almost negligible for parallel 
polarization but lead to a strong depolarization in perpendicular
direction: The C(3,3) tube is almost transparent
up to 5 eV \cite{tolis03} in agreement with the experimental observation
\cite{li01}. 

In conclusion, the calculations for optical absorption of small C and
BN-nanotubes display a variety of features beyond the excitation
between first and second Van-Hove singularities.
In C-tubes the position of the first absorption peak strongly varies
with the tube indices while in BN-tubes the first peak is determined
by the band gap of BN and is therefore mostly independent of $(n,m)$. 
Some of the fine-structure which distinguishes BN-tubes 
and C-tubes of different chirality is only visible in the UV
region which gives rise to the hope that this energy regime will
be probed in the future.
Still, features related to electron-hole attraction (excitonic effects)
are missing in the calculations. They can play a role in both C and BN tubes, 
leading to new structure in the band-gap and a redistribution of oscillator 
strength. They may explain the anomalous $E_{11}/E_{22}$
ratio \cite{bac03} measured recently for C-tubes \cite{contribs}.
Work along these lines for BN-tubes is in progress.


\bibliographystyle{aipproc}   

\begin{thebibliography}{1}
\bibitem{noteps}
The dielectric function $\epsilon_M(\omega)$ is defined in 
the context of 3-dim. periodic systems. 
Isolated tubes have only a 1-dimensional translational symmetry.
However, the use of a plane-wave expansion requires effectively the
use of a three-dimensional super-cell. Therefore, we perform calculations
for a periodic array of nanotubes with an intertube distance of 10 a.u.
This keeps the interaction between neighboring tubes low and renders
nearly isolated tubes. Analogously, for the 2-dim. ``single'' sheets of 
C and BN, we perform calculations for an infinite stacking of sheets 
with an inter-sheet distance of 12 a.u. The absolute value of $\epsilon_M(
\omega)$ scales with the dimension of the employed super-cell, but 
we are interested only in the relative absorption cross section
$\sigma(\omega)$ in arbitrary units.
\bibitem{params}
The cutoff energy of the plane-wave basis is 50 Ry.
Core-electrons are simulated by Troullier-Martins pseudopotentials. 
Calculations are performed in the local-density approximation (LDA).
\bibitem{nvalues}
For all systems, we compute bands up to $N=4n_{occ.}$, where $n_{occ.}$ is
the number of occupied bands.
\bibitem{kvalues}
For the 2-dim. Brillouin zone of the
C-sheet, we use a sampling of $87 \times 87$ $k$-points,
for the BN-sheet $67 \times 67$. For the 1-dim. Brillouin zones of the
C-tubes, we use of $44$ $k$-points, for the BN-tubes $20$.
\end{thebibliography}

\end{document}